\documentclass[fleqn,usenatbib]{mnras}

\usepackage[T1]{fontenc}
\usepackage{ae,aecompl}

\usepackage{graphicx}	
\usepackage{amsmath}	
\usepackage{amssymb}	

\usepackage{subfig} 

\title[Desorption of NH$_3$ from Forsterite]{Thermal Desorption of Ammonia from Crystalline Forsterite Surfaces}

\author[T. Suhasaria et al.]{
T. Suhasaria,
J. D. Thrower,\thanks{E-mail: john.thrower@uni-muenster.de}
H. Zacharias, \thanks{E-mail: H.Zacharias@uni-muenster.de}
\\
Physikalisches Institut, Westf{\"a}lische Wilhelms-Universit{\"a}t, 48149 M{\"u}nster, Germany
}

\date{Accepted 2015 September 21. Received 2015 September 9; in original form 2015 July 22}

\pubyear{2015}

\begin{document}
\label{firstpage}
\pagerange{\pageref{firstpage}--\pageref{lastpage}}
\maketitle

\begin{abstract}
The thermal desorption of ammonia (NH$_3$) from single crystal forsterite (010) has been investigated using temperature-programmed desorption. The effect of defects on the desorption process has been probed by the use of a rough cut forsterite surface prepared from the cleaved forsterite sample. Several approaches have been used to extract the desorption energy and pre-exponential factor describing the desorption kinetics. In the sub-monolayer coverage regime, the NH$_3$ desorption shows a broad distribution of desorption energies, indicating the presence of different adsorption sites, which results in an apparent coverage-dependent desorption energy. This distribution is sensitive to the surface roughness with the cut forsterite surface displaying a significantly broader distribution of desorption energies compared to the cleaved forsterite surface. The cut forsterite surface exhibits sites with desorption energies up to 62.5 kJ mol$^{-1} $ in comparison to a desorption energy of up to 58.0 kJ mol$^{-1} $ for the cleaved surface. Multilayer desorption is independent of the nature of the forsterite surface used, with a desorption energy of ($25.8\pm0.9$) kJ mol$^{-1} $. On astrophysically relevant heating time-scales, the presence of a coverage dependent desorption energy distribution results in a lengthening of the NH$_3$ desorption time-scale by $5.9\times 10^4$ yr compared to that expected for a single desorption energy. In addition, the presence of a larger number of high-energy adsorption sites on the rougher cut forsterite surface leads to a further lengthening of \textit{ca.} 7000 yr.
\end{abstract}




\section{Introduction}
Dust grain surfaces play a key role in molecule formation in the interstellar medium (ISM; \citep{Williams2002,Herbst2005}). Over the last few decades, observations of infrared emission and absorption spectroscopic features of dust grains have revealed insights into their composition \citep{Draine2003, Williams2005}. These dust grains are thought to be mainly silicate or carbonaceous in nature \citep{Snow1996}. The \textit{Infrared Space Observatory} (ISO) mission identified the presence of silicates based on the 9.7 and 18-$\mu$m IR emission and absorption bands \citep{Waters2002}. It is found that silicates in the ISM are predominantly amorphous \citep{Li2002}. On the other hand, in the circumstellar regions of pre- and post-main sequence stars, up to 20\% of silicates are present in crystalline form \citep{Molster2005} as observed using the short wavelength spectrometer on board ISO \citep{Waters1996}. Previous measurements \citep{Harker2002,Gail2004} indicate that crystalline silicates can be formed only by high-temperature annealing which is consistent with their higher abundance in the vicinity of evolved stars. After their formation, grains are injected in the ISM, being driven by the stellar winds. Olivine (Mg$_{2x}$Fe$_{2-2x}$SiO$_{4}$) is an abundant class of silicate minerals that constitutes a large fraction of interstellar dust grains \citep{Campins1989}. Forsterite (Mg$_{2}$SiO$_{4}$), the magnesium-rich end-member of the olivine family, is one of the most abundant crystalline silicates \citep{Tielens1998,Molster2002} in dust clouds near young stellar objects and can be considered as a convenient dust grain model to understand the interaction between interstellar molecules and crystalline silicate grain surfaces.

Observations have revealed that in cold, dense regions of the ISM, for example within dark clouds, the dust grains are coated with layers of ice consisting of various small molecules including H$_{2}$O, CO, CO$_{2}$, NH$_{3}$, CH$_{3}$OH, with H$_{2}$O ice dominating \citep{Whittet1998}. H$_{2}$ along with H$_{2}$O and many other molecules are thought to form predominately \textit{in situ} through grain surface reactions \citep{Hollenbach1971,Tielens1982,Ioppolo2008}. This also applies for ammonia, NH$_{3}$, for which gas phase reactions alone are insufficient to explain the observed abundance \citep{Pauls1983}. NH$_{3}$ is observed to be present as a relatively high abundance constituent of interstellar ices, showing an abundance of up to 30\% with respect to water ice \citep{Whittet1996,Lacy1998,Chiar2000,Dartois2001,Gibb2001}. NH$_3$ is an important carrier of interstellar nitrogen and therefore plays a central role in the nitrogen chemistry of the ISM \citep{Tielens1982, Nejad1990,Charnley2002}. Solid phase NH$_{3}$ has also been identified on surfaces of several icy bodies including Charon \citep{Brown2000}, Miranda \citep{Bauer2002}, the Kuiper belt object Quaoar \citep{Jewitt2004} and Enceladus \citep{Emery2005}, and so laboratory studies of NH$_{3}$ ice are of interest not only in the context of the ISM, but also the outer Solar system \citep{Moore2007}.

Thermal desorption is an important mechanism by which molecules within the icy mantles on the dust grain are returned to the gas phase \citep{Roberts2007}. It occurs within star-forming regions when the dust grains are heated and the icy mantles are sublimated \citep{VanDishoeck1998}. Temperature-programmed desorption (TPD) measurements performed in the laboratory replicate the sublimation of icy mantles but over a significantly shorter time-scale. Desorption kinetic parameters obtained from such measurements employing representative model grain-ice systems can be used in astrophysical models to gain more insight into the role played by desorption processes on astrophysically relevant time-scales. More recently, the contribution of gas-grain interactions and reactions within ice films has been explored by employing the desorption kinetics obtained from TPD measurements to describe the thermal desorption of ices in three-phase astrochemical models \citep{Garrod2013}.

NH$_{3}$ desorption kinetics have been studied in great detail for single crystal metal surfaces including Ru(0001) \citep{Benndorf1983}, Au(111) \citep{Kay1989}, Al(111) \citep{Kim1997} and Ag(111) \citep{Szulczewski1997,Szulczewski1998}. In order to understand the interaction between molecules and interstellar grain surfaces, as well as determine the desorption kinetics, the use of appropriate grain mimics based on carbonaceous or siliceous materials is essential. The desorption of NH$_{3}$ from the highly oriented pyrolytic graphite (HOPG) surface was observed, through TPD measurements, to follow fractional order desorption kinetics with a desorption energy of $E_{\mathrm{des}}=23.2$ kJ mol$^{-1}$ for multilayers \citep{Bolina2005}. NH$_{3}$ desorption from silicate surfaces has also been investigated by considering the quartz(0001) surface \citep{Grecea2011}. Zero order kinetics, attributed to multilayer desorption, as well as two additional desorption features attributed to monolayer (ML) desorption from (0001) terraces and, for the lowest coverages, surface defects, respectively, were observed. NH$_{3}$ desorption from amorphous silicate thin films grown on a gold plated copper surface shows a desorption energy distribution in the range of \textit{ca.} 23.3-28.3 kJ mol$^{-1}$ \citep{He2015}. Quartz crystal microbalance techniques have also been used to investigate the sublimation of multilayer NH$_{3}$ ice, yielding a sublimation energy of 31.8 kJ mol$^{-1}$ \citep{Luna2014}. In the case of silicate minerals incorporating one or other metal atoms, the resulting increase in potential adsorption sites would be expected to influence the adsorption energies of bound species. We have investigated both this effect, and the impact of surface roughness on the kinetics for NH$_3$ desorption from forsterite-derived surfaces.

We have developed two silicate grain mimics derived from single crystal forsterite, the stable (010) surface plane exposed through mechanical cleavage and a rougher, defected surface produced by cutting the crystal. The cleaved surface allows us to assess the adsorption on the smooth crystalline surface itself, whereas the cut surface provides a more representative surface due to presence of a higher density of defects as expected to be present in astrophysical environments. The forsterite(010) crystal has been investigated extensively in recent computational studies as an interstellar dust grain analogue to study the interaction of H atoms \citep{Downing2013,Garcia-Gil2013}, the surface catalysed formation of H$_2$ \citep{Goumans2009} and the adsorption and surface formation of H$_2$O \citep{DeLeeuw2000,Muralidharan2008,Goumans2009,Asaduzzaman2013}. Only recently has crystalline olivine been explored in experimental studies of the desorption kinetics of simple ice species. Smith and co-workers investigated the desorption kinetics of CO$_{2}$ and H$_{2}$O from the natural crystalline olivine (011) surface, with the silicate incorporating both Mg$^{2+}$ and Fe$^{2+}$ \citep{Smith2014} and obtained the desorption energy and pre-exponential factor describing the desorption process.

Using TPD measurements we derive the kinetic parameters for NH$_{3}$ desorption from the cleaved and cut forsterite (Mg$_2$SiO$_4$) surfaces following adsorption at 78 K. At this temperature crystalline solid phase NH$_3$ is obtained. Although this is higher than the grain surface temperatures (10-15 K) expected under typical dense cloud conditions, this nevertheless allows us to investigate the temperature range over which desorption occurs. Although NH$_{3}$ ices generally do not exist in pure form in the ISM, it is essential to study such simple systems to understand the interactions on different model interstellar grain surfaces. Ice mixtures result in additional complexity, for example \citet{Martin2014} investigated the thermal desorption of NH$_{3}$ from various ice mixtures containing H$_{2}$O, CO, CO$_{2}$, CH$_{3}$OH and NH$_{3}$. The presence of these additional components, particularly H$_2$O and CH$_3$OH, leads to trapping/co-desorption behaviour, with the bulk of the NH$_3$ desorbing at higher temperature along with these other components. So-called molecular volcano desorption, occurring during the crystallization of the H$_2$O and CH$_3$OH ice was also observed, in contrast to previous measurements by \citet{Collings2004} who also observed the co-desorption of NH$_3$ with H$_2$O from an NH$_{3}$/H$_{2}$O mixed ice but no molecular volcano. In this work we focus primarily on the interaction between NH$_3$ and our model silicate grain surfaces.



\section{Experimental}

\subsection{Sample Preparation}

\begin{figure}
\includegraphics[width=\columnwidth]{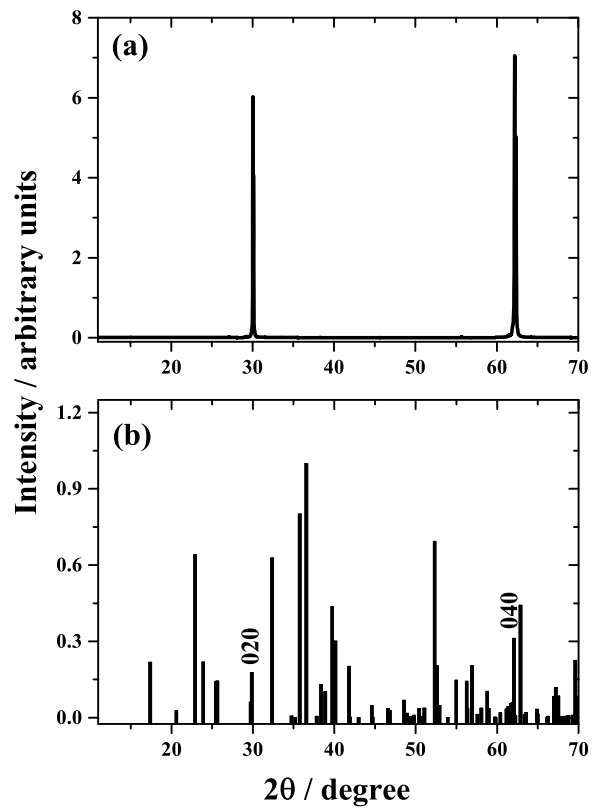}
\caption{Reflective X-ray diffraction pattern for the cleaved forsterite sample in the range of $2\theta$ = $10^{\circ}$ to $70^{\circ}$. Measurements were performed using Cu K$_\alpha$ radiation. (a) Experimental diffraction data; (b) reference data (\citet{Kirfel2005}). The strongest reflections observed at 2$\theta$ = $30.04^{\circ}$ and 2$\theta$ = $62.18^{\circ}$ correspond to the (020) and (040) reflections respectively, confirming the dominance of the (010) surface plane. Note that the values on the intensity axes are not directly comparable between the two subfigures.}
\label{Fig:XRD}
\end{figure}

The silicate grain mimics employed in the thermal desorption studies were prepared from a synthetic forsterite single crystal (Oxide Corporation) which was characterized using IR reflectance measurements and X-ray photoelectron spectroscopy (XPS). The reflectance spectra show strong reflections in the $\tilde\nu\sim$ 900-1000 cm$^{-1}$ region along all three crystal axes indicative of the Si-O stretching and bending modes, in good agreement with previous measurements \citep{Sogawa2006,Suto2006}. The XPS spectra show that Si, Mg and O atoms are present on the substrate surface. No features attributable to Fe were observed as expected for the case of forsterite as opposed to an Fe containing member of the olivine family.  The dimension of this single crystal was $10\times10\times10$ mm$^{3}$ and it was subsequently mechanically cleaved in the centre to expose the most stable (010) surface plane of forsterite \citep{King2010}; the cleaved crystal was 4 mm thick. The crystal surface orientation was confirmed by reflective diffraction measurements performed with Cu K$\alpha$ radiation employing a Bruker D8 Advance diffractometer. The diffraction spectra are presented and compared with the reported values \citep{Kirfel2005} in Figure~\ref{Fig:XRD}. The diffraction peaks at $2\theta=30.04^{\circ}$ and $2\theta=62.18^{\circ}$ are in good agreement with those expected for the (020) and (040) reflections, respectively \citep{Kirfel2005}. This demonstrates that the (010) surface plane dominates the cleaved forsterite sample. To prepare the rough sample, the other half of the cleaved crystal was cut parallel to the exposed (010) plane using a precision vertical diamond wire saw with a 300 $\mu$m wire diameter (Well 3242), resulting in a 1 mm thick sample. Reflective diffraction measurements again revealed a dominance of the (010) crystal plane. However, compared to cleaving the crystal, such a procedure will lead to an atomically rough surface exhibiting a large density of defect features.


\subsection{Thermal Desorption Measurements}

\begin{figure}
\includegraphics[width=\columnwidth]{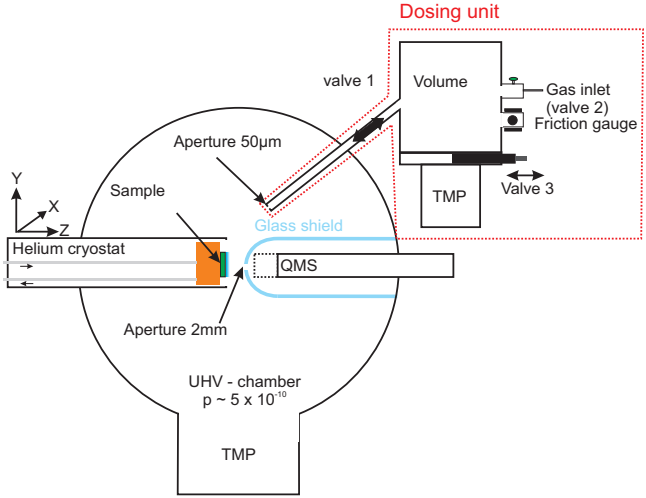}
\caption{(Colour online) Schematic of the experimental set-up for temperature programmed desorption measurements. The red dotted line highlights the dosing unit.}
\label{Fig:Set-up}
\end{figure}

TPD measurements were performed in an ultra-high vacuum (UHV) chamber pumped by a turbomolecular pump to a base pressure of better than $5\times10^{-10}$ mbar after bake-out. The set-up is shown schematically in Figure~\ref{Fig:Set-up}. The chamber is equipped with a flowing liquid nitrogen cryostat (Cryovac Konti UHV) mounted on an x-y-z-$\theta$ manipulator. The sample is clamped to the end of the cryostat with silver paste being used to ensure good thermal contact to the sample holder. The sample can be locally heated using a 50 W halogen bulb situated behind the sample holder. The sample temperature is monitored using a K-type (chromel-alumel) thermocouple which is pressed between the front face of the sample and the securing clamp. Prior to each experiment, the sample was heated to 465 K in UHV for several minutes to desorb any volatile adsorbates and then held at 200 K, above the sublimation temperature of water, until the pressure fell below $10^{-9}$ mbar. The sample was then rapidly cooled to the base temperature of T$_s\sim78$ K prior to exposure to NH$_{3}$.

Ammonia (99.999\% ; Westfalen AG) was deposited on to the sample through a dosing system (see Figure~\ref{Fig:Set-up}). This consists of a small high vacuum chamber with a well-defined volume which is pumped by a separate turbomolecular pump from which it can be isolated by means of a gate valve (valve 3 in Figure~\ref{Fig:Set-up}). The pressure in the dosing chamber is measured using a calibrated spinning rotor gauge (MKS instruments). Gases can be introduced into the dosing chamber \textit{via} a precision leak valve (valve 2 in Figure~\ref{Fig:Set-up}). The dosing system can be isolated from the main chamber by means of a second gate valve (valve 1 in Figure~\ref{Fig:Set-up}). A tube terminated with a 50 $\mu$m aperture extends into the main chamber in order to constrain the exposure to the sample region of the cryostat. Initially, the valves (1-3) of the dosing unit are closed and the defined volume is then filled with the gas of interest to a particular pressure. Valve (1) is then opened and the gas is dosed on to the sample which is positioned \textit{ca.} 5 mm from the aperture. After a defined dosing time valve (1) is closed. The exposure is increased by increasing the dosing time. During exposure only a small pressure rise to about $5.5\times10^{-10}$ mbar in the main chamber is observed even during multilayer dosing, confirming the localized nature of the dosing.

Desorbing molecules were detected using a quadrupole mass spectrometer (QMS; Hiden Analytical HAL 7). NH$_{3}$ desorption was monitored primarily through its parent ion at $m/z=17$. The NH$_{2}^{+}$ fragment at $m/z=16$ showed the same behaviour, and the non-detection of any desorption signal at $m/z=18$ confirmed the absence of H$_{2}$O adsorption on the sample. The QMS is contained within a glass shield with an entrance aperture of 2 mm. This reduces the detection of molecules desorbing from surfaces other than that of the sample of interest, further constraining the desorption signal to that corresponding to the forsterite crystal \citep{Feulner1980}. Desorption experiments were carried using a linear heating ramp of $\mathrm{d}T/\mathrm{d}t=(0.3\pm {0.01})$ K s$^{-1}$ using a proportional-integral-derivative controller (PID; Funktechnik-Elektronik TRM 1). The pre-amplified thermocouple voltage from the PID controller is then
recorded by the QMS \textsc{massoft professional} software package simultaneously
with the ion signal.

\section{Results \& Analysis}

\subsection{\texorpdfstring{Thermal Desorption of NH$_3$}{Thermal Desorption of NH3}}

\begin{figure}
\includegraphics[width=\columnwidth]{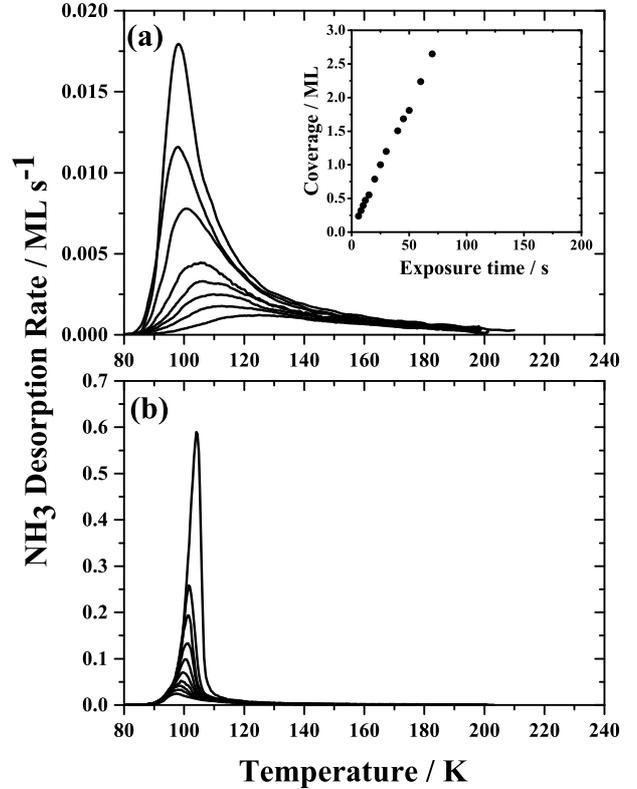}
\caption{TPD traces for NH$_3$ deposited on cleaved forsterite(010) at 78 K and heated at 0.3 K s$^{-1}$ for increasing initial NH$_3$ coverages of (a) $\Theta_{\mathrm{i}}$ = 0.2, 0.3, 0.4, 0.5, 0.6, 0.8, 1.0 and 1.2 ML and (b) increasing multilayer coverages of $\Theta_{\mathrm{i}}$ = 1.3, 1.5, 1.7, 1.8, 2.2, 2.7, 3.2, 4.0, 4.8 and 9.4 ML.  The inset in (a) shows the surface coverage in ML obtained with the ammonia exposure time in seconds. All traces have been smoothed using an adjacent averaging procedure.}
\label{Fig:TPD_Cleaved}
\end{figure}

Figure~\ref{Fig:TPD_Cleaved} shows NH$_{3}$ TPD traces for desorption from the cleaved forsterite(010) surface following deposition at 78 K. We denote coverage in ML as $\Theta$ and define $\Theta_{\mathrm{i}}$ as the initial coverage at the start of the TPD measurement that results from the deposition. Figure~\ref{Fig:TPD_Cleaved}(a) shows the desorption traces resulting from the smallest exposures of NH$_{3}$ which result in coverages ranging from $\Theta_{\mathrm{i}}= 0.2$ to 1.2 ML, revealing the gradual development of a peak which we attribute to desorption from the first ML of NH$_{3}$ on the forsterite surface. We define the ML peak with $\Theta_{\mathrm{i}}=1$ ML as the coverage at which the leading edges of the TPD curves become coincident, a feature associated with the onset of zero order desorption kinetics from multilayer films. For a given exposure, the relative NH$_{3}$ coverage in ML was obtained from the total integrated TPD yield relative to that of the first ML. It is important to note that the absolute surface coverage and therefore exposure required to complete the first ML depends on the number of available adsorption sites. Thus, the definition of a ML is not necessarily equivalent for our two surfaces. To highlight this we report in the inset to Figure~\ref{Fig:TPD_Cleaved}(a) the coverage in ML achieved for different dosing times which gives a measure of the exposure (\textit{ca.} 25 s) required to achieve a full ML for the crystalline forsterite surface.  For sub-ML coverages, the peak maximum shifts from 127 K at $\Theta_{\mathrm{i}}=0.2$ ML to 98 K at 1.0 ML. In all cases, the sub-ML desorption traces show a high temperature tail extending to temperatures of \textit{ca.} 210 K. This is attributed to desorption of NH$_{3}$ molecules that are bound to the forsterite surface in a range of adsorption sites. Figure~\ref{Fig:TPD_Cleaved}(b) shows TPD traces for coverages of $\Theta_{\mathrm{i}}=1.3$ ML to 9.4 ML. We observe a peak maximum at around $T_{\mathrm{max}}=100$ K, shifting to higher temperature with increasing coverage as expected for multilayer desorption exhibiting zero order kinetics, typical for molecular films with attractive adsorbate-adsorbate interactions.

\begin{figure}
\includegraphics[width=\columnwidth]{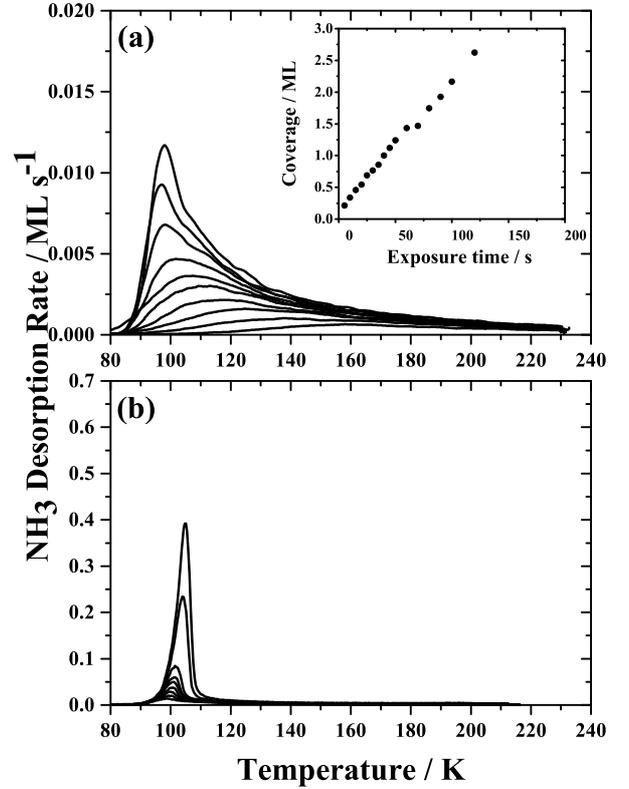}
\caption{TPD traces for NH$_3$ deposited on cut forsterite at 78 K and heated at 0.3 K s$^{-1}$ for increasing initial NH$_3$ coverages of (a) $\Theta_{\mathrm{i}}$ = 0.2, 0.3, 0.5, 0.6, 0.7, 0.8, 0.9, 1.00, 1.1 and 1.2 ML and (b) increasing multilayer coverages of $\Theta_{\mathrm{i}}$ = 1.3, 1.4, 1.5, 1.8, 1.9, 2.2, 2.6, 4.6 and 8.3 ML. The inset in (a) shows the surface coverage in ML obtained with the ammonia exposure time in seconds. All traces have been smoothed using an adjacent averaging procedure.}
\label{Fig:TPD_Cut}
\end{figure}

TPD traces for NH$_{3}$ desorption from the rougher, cut forsterite surface following deposition at 78 K are shown in Figure~\ref{Fig:TPD_Cut}. NH$_{3}$ coverages were determined in the same way as for the cleaved forsterite surface. The inset to Figure~\ref{Fig:TPD_Cut} (a) again shows the coverage in ML achieved for increasing dosing times, which for the rough cut forsterite surface gives a necessary exposure time of \textit{ca.} 40 s to complete the ML. This confirms that the rough surface does indeed present a higher number of adsorption sites compared to the cleaved surface. Assuming equivalent dosing conditions, this suggests that the number of adsorption sites on the rough surface is approximately 1.6 times that on the cleaved surface. TPD traces for initial coverages of $\Theta_{\mathrm{i}}=0.2$-1.2 ML are shown in Figure~\ref{Fig:TPD_Cut}(a). The high temperature tail, which is a characteristic feature of NH$_{3}$ desorption from both surfaces, now extends to higher temperatures in excess of 230 K. For an initial coverage of $\Theta_{\mathrm{i}}=0.2$ ML a broad peak is observed with maximum intensity at around $T_{\mathrm{max}}$ = 157 K. At saturation coverage, the ML peak is observed at around $T_{\mathrm{max}}=98$ K, as was the case for 1 ML NH$_{3}$ desorption from cleaved forsterite(010). The increase in desorption rate at higher temperatures compared to the cleaved sample is consistent with the presence of a higher density of defects (\textit{e.g.} steps, kinks) on the cut surface, which provide higher energy adsorption sites. Figure~\ref{Fig:TPD_Cut}(b) shows NH$_{3}$ desorption traces from $\Theta_{\mathrm{i}}= 1.3$ to 8.3 ML. This peak again shifts to higher temperature with increasing coverage as in the case of cleaved forsterite, with peak centred initially at around 100 K. The multilayer desorption traces for NH$_{3}$ adsorbed on cut forsterite appear very similar to those obtained for NH$_{3}$ from cleaved forsterite as shown in Figure~\ref{Fig:TPD_Cleaved}(b), consistent with a decreasing influence of the underlying substrate on desorption from multilayer films.

\subsection{Desorption Energy Determination}

In order to compare the desorption behaviour between the two substrates it is necessary to extract from the experimental desorption traces the underlying kinetic parameters. Thermal desorption can be described using the Polanyi-Wigner equation \citep{King1975,DeJong1990}:


\begin{equation}
r_{\mathrm{des}} = -\frac{\mathrm{d}\Theta}{\mathrm{d}t}(\Theta,t) = \nu_n \Theta^n\exp\left[-\frac {E_{\mathrm{des}}}{RT}\right],
\label{eq:PW eq}
\end{equation}

which yields the desorption rate, $r_{\mathrm{des}}$, as the rate of change of coverage $\Theta$ with respect to time, $t$. $\nu_n$ denotes the pre-exponential factor associated with the desorption order, $n$, $E_{\mathrm{des}}$ the desorption energy, $R$ the universal gas constant, and $T$ the surface temperature. Since physisorption of small molecules like NH$_{3}$ is typically a non-activated process, the desorption energy can be considered equal to the adsorption energy of the adsorbate to the surface, \textit{i.e.}, $E_{\mathrm{des}}=E_{\mathrm{ads}}$.

For multilayer desorption the coincident leading edge behaviour observed in the TPD traces suggests zero order desorption kinetics. The desorption energies were then determined using an Arrhenius analysis of the leading edge region as described previously \citep{Habenschaden1984,DeJong1990}. The Polyani-Wigner equation for the zero order ($n=0$) case can be rearranged as follows:

\begin{equation}
\ln (r_{\mathrm{des}}) = \ln\left(-\frac {\mathrm{d}\Theta}{\mathrm{d}t}\right) = \ln (\nu_0) -\frac {E_{\mathrm{des}}}{R}\left[\frac {1}{T}\right].
\label{eq:PW(Zero order eq}
\end{equation}

\begin{figure}
\includegraphics[width=\columnwidth]{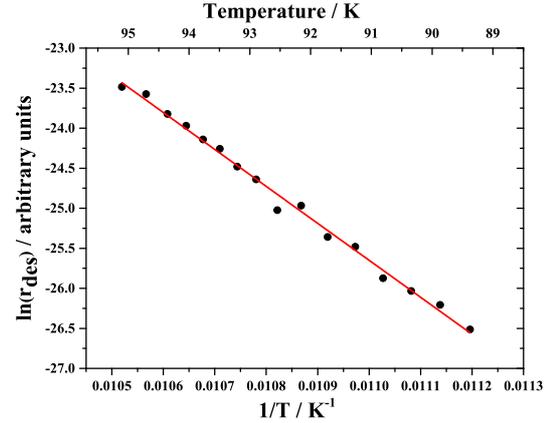}
\caption{(Colour online) Arrhenius plot of NH$_3$ desorption for an initial coverage of $\Theta_{\mathrm{i}}=1.7$ ML from a cleaved forsterite(010). The black filled circles are the experimental data points while the solid red line is a linear least square fit to the data. From the slope of this fit a desorption energy of $E_{\mathrm{des}}=(25.1\pm1.0)$ kJ mol$^{-1}$ is obtained.}
\label{Fig:ArrheniusPlot}
\end{figure}

The TPD traces can then be plotted in Arrhenius form with $\ln(r_{\mathrm{des}})$ \textit{versus} $1/T$ for which a linear fit to the slope yields the desorption energy. A typical Arrhenius plot, obtained for the desorption of NH$_{3}$ from the cleaved forsterite (010) surface at an initial coverage of $\Theta_{\mathrm{i}}= 1.7$ ML is shown in Figure~\ref{Fig:ArrheniusPlot}. For both surfaces and several initial coverages in the multilayer coverage regime, Arrhenius plots were constructed, from which a mean value of $E_{\mathrm{des}}$ was derived. The leading edge method yields effectively identical desorption energies of $E_{\mathrm{des}}=(25.8\pm0.9)$ kJ mol$^{-1}$ and $(25.8\pm0.7)$ kJ mol$^{-1}$ for NH$_{3}$ mutlilayer desorption from cleaved and cut forsterite, respectively.

As a first approximation, Redhead's peak maximum method \citep{Redhead1962} was used to obtain the desorption energy in the sub-ML coverage regime. This method assumes first order desorption kinetics and a value of the pre-exponential of $\nu_1=10^{13}$ s$^{-1}$ for small, weakly bound adsorbates. First order desorption from the sub-ML coverage regime generally shows a characteristic asymmetric peak which remains at a fixed temperature with increasing coverage for cases where the kinetic parameters $\nu_n$, $E_{\mathrm{des}}$ and $n$ are independent of $\Theta$ and $T$. In the present case we see that the peak shifts to lower temperatures with increasing coverage, suggesting that the kinetics are coverage dependent. Employing this approach for several initial coverages thus serves to provide insight into this dependence. For a given TPD trace corresponding to a particular initial coverage, the Redhead desorption energy is given by:

\begin{equation}
E_{\mathrm{des}} = RT_{\mathrm{max}}\left[\ln \left(\frac {\nu T_{\mathrm{max}}} {\beta}\right)-3.46\right],
\label{eq:Redhead eq}
\end{equation}

where $T_{\mathrm{max}}$ denotes the temperature at which the maximum desorption rate is observed and $\beta$ is the heating rate. The error in the determination of the desorption energy is obtained based on that in the value of $T_{\mathrm{max}}$ which we estimate at $\pm 1$ K and by considering an order of magnitude variation in $\nu_1$ from $10^{12}$ to $10^{14}$ s$^{-1}$, the typical range expected for small, weakly bound adsorbates. On cleaved forsterite for an initial NH$_{3}$ coverage of $\Theta_{\mathrm{i}}=0.2$ ML this yields a desorption energy of $E_{\mathrm{des}}=(34.1\pm2.7)$ kJ mol$^{-1}$. For the cut forsterite surface and a similar initial coverage of $\Theta_{\mathrm{i}}=0.2$ ML the value of $E_{\mathrm{des}}=(42.5\pm3.3)$ kJ mol$^{-1}$ obtained is significantly higher, reflecting the presence of higher energy adsorption sites on the rougher cut surface compared to the cleaved (010) surface. Table~\ref{Table:Redhead} shows the desorption energies obtained for both surfaces at all the initial coverage values deposited up to $\Theta_{\mathrm{i}}$ = 1 ML. It is evident that the difference between the desorption energies obtained for the two surfaces is significant only at the lowest coverages. As the initial coverage increases the desorption energy values of NH$_{3}$ on the two surfaces tend to become similar. This suggests the importance of adsorbate-substrate interaction over adsorbate-adsorbate interaction at low coverages.

\begin{table}
\centering
\caption{Desorption energies determined by Redhead's peak maximum method. \label{Table:Redhead}}
\begin{tabular}{ccc}
 \hline
 \multicolumn{1}{c}{}  & \multicolumn{1}{c}{Cleaved forsterite(010)}  & \multicolumn {1}{c}{Cut forsterite}\\
\hline
  \text{Coverage($\Theta$)} & \text{$E_{\mathrm{des}}$} & \text{$E_{\mathrm{des}}$}\\ (ML) & (kJ mol$^{-1}$) & (kJ mol$^{-1}$) \\
 \hline
0.2 & 34.1$\pm2.7$ & 42.5$\pm3.3$ \\
0.3 & 31.0$\pm2.5$ & 37.7$\pm2.9$ \\
0.4 & 30.0$\pm2.4$ & - \\
0.5 & 29.4$\pm2.4$ & 34.1$\pm2.7$ \\
0.6 & 28.2$\pm2.3$ & 31.5$\pm2.5$ \\
0.7 & - & 29.7$\pm2.4$ \\
0.8 & 26.7$\pm2.2$ & 28.4$\pm2.3$ \\
0.9 & - & 26.0$\pm2.1$ \\
1.0 & 26.5$\pm2.2$ & 27.7$\pm2.2$ \\
\hline
\end{tabular}
\end{table}

As we have discussed, Redhead's peak maximum method does not fully account for the experimentally observed desorption traces in particular the high temperature tail that is evident for desorption from both surfaces. This strongly suggests coverage dependent kinetic parameters. The representative desorption energy values provided by the Redhead method indicate the presence of higher energy adsorption sites that are initially occupied at small exposures and subsequent population of lower binding energy adsorption sites with increasing exposure. To understand the sub-ML desorption kinetics in more detail an expression for the coverage dependent desorption energy, $E_{\mathrm{des}}(\Theta)$ was obtained by inverting the Polanyi-Wigner equation:

\begin{equation}
E_{\mathrm{des}}(\Theta) = -RT \ln \left[- \frac {\mathrm{d}\Theta /\mathrm{d}t}{\nu_1 \Theta}\right].
\label{eq:invertedPW}
\end{equation}

This procedure has been employed previously for other systems that display a coverage dependent binding energy, \textit{e.g.} CO on MgO(100)\citep{Dohnalek2001}, D$_2$ on amorphous solid water \citep{Hornekaer2005,Amiaud2007}, N$_2$ on amorphous solid water \citep{Zubkov2007}, benzene (C$_6$H$_6$) on amorphous silica \citep{Thrower2009} and NH$_3$ on vapour deposited amorphous silicate \citep{He2015}.

Desorption energies were then calculated for the individual TPD traces as a function of the changing coverage assuming, as in the Redhead method, a pre-exponential value $\nu_1=10^{13}$ s$^{-1}$. This analysis was performed for all initial coverages up to 1 ML for both the cleaved and cut forsterite surfaces. Figure~\ref{Fig:Distribution} shows the resulting desorption energy curves as a function of coverage obtained using Equation~\ref{eq:invertedPW} for an initial coverage of $\Theta_{\mathrm{i}}=1$ ML for the cleaved and cut forsterite surfaces. For other initial coverages (data not shown), the $E_{\mathrm{des}}(\Theta)$ curves obtained are coincident with those obtained for an initial coverage of $\Theta_{\mathrm{i}}=1$ ML over the corresponding coverage range. This clearly indicates the coverage dependent energy distribution at low exposures. For inclusion in kinetic simulations the $E_{\mathrm{des}}(\Theta)$ curve for $\Theta_{\mathrm{i}}=1$ ML was described by a first order exponential fit which gives the full coverage dependence of the desorption energy in the sub-ML regime, as shown in Figures~\ref{Fig:Distribution} (a) and (b). The corresponding distribution of desorption energies $P(E_{\mathrm{des}})$ was then obtained from this fit using the following expression \citep{Zubkov2007}:

\begin{figure}
\includegraphics[width=\columnwidth]{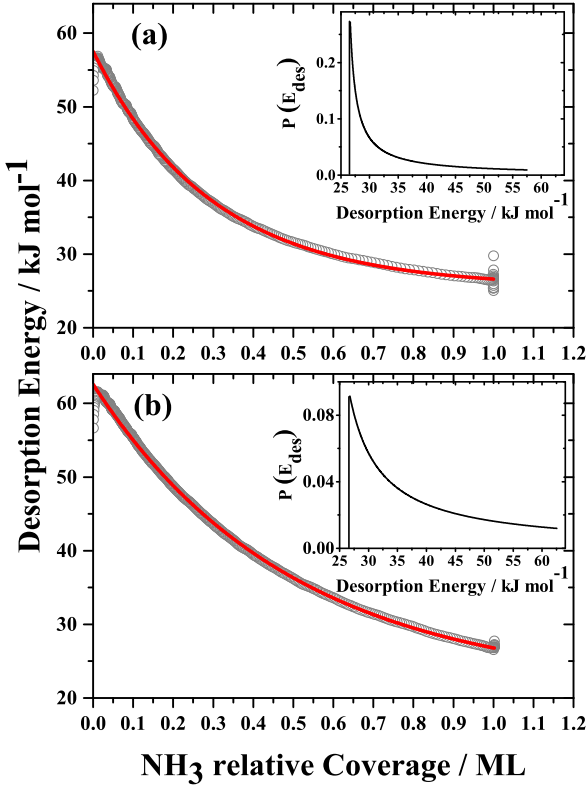}
 \caption{(Colour online) Desorption energy as a function of coverage ($E_{\mathrm{des}}(\Theta)$) as obtained from the TPD traces for 1 ML of NH$_3$ adsorbed on (a) cleaved and (b) cut forsterite. The open circles are the result of directly inverting the experimental traces while the solid lines are single component exponential fits to the data. The insets show the associated desorption energy distributions.}
\label{Fig:Distribution}
\end{figure}

\begin{equation}
P(E_{\mathrm{des}}) = -\frac{\mathrm{d}\Theta}{\mathrm{d}E_{\mathrm{des}}}.
\label{eq:Distribution eq}
\end{equation}

This distribution of desorption energies for the cleaved forsterite surface is shown as inset in Figure~\ref{Fig:Distribution}(a) and clearly demonstrates that, although the desorption energy is dominated by values in the $26-30$ kJ mol$^{-1}$ range, sites displaying desorption energies of up to $E_{\mathrm{des}}=58$ kJ mol$^{-1}$ are also present on the cleaved forsterite(010) surface. The same approach to extract the distribution of desorption energies was performed for the desorption of NH$_{3}$ from the cut forsterite surface as shown in the inset of Figure~\ref{Fig:Distribution}(b). There is a clear difference between the two surfaces in the observed distribution of desorption energies. Higher desorption energies of up to $E_{\mathrm{des}}\sim62.5$ kJ mol$^{-1}$ are obtained for the cut surface compared to the maximum desorption energy of $E_{\mathrm{des}}\sim 58$ kJ mol$^{-1}$ for the cleaved forsterite surface.  It is found that \textit{ca.} $77\%$ of adsorption sites yield desorption energies greater than 30 kJ mol$^{-1}$ for cut forsterite surface compared to \textit{ca.} $58\%$ of sites in the case of cleaved forsterite. For the cut forsterite surface, a higher maximum desorption energy and a generally higher population of high-energy sites compared to the cleaved forsterite(010) surface is observed. This is consistent with the expected higher density of defects on this surface that can present stronger adsorption sites for adsorbate molecules.

\begin{figure}
\includegraphics[width=\columnwidth]{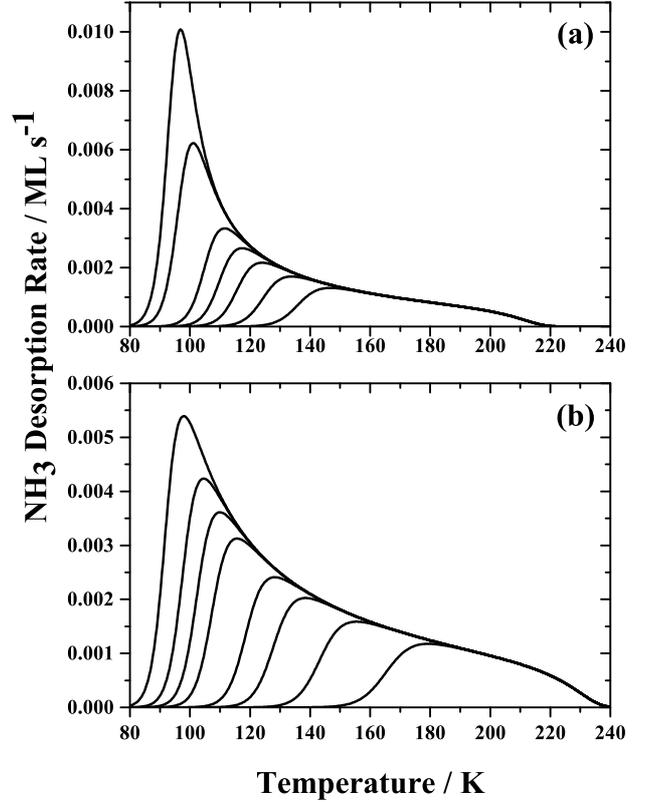}
\caption{Simulated sub-ML NH$_3$ TPD traces for (a) cleaved fosterite with $\Theta_{\mathrm{i}}$ = 0.2, 0.3, 0.4, 0.5, 0.6, 0.8 and 1.00 ML and (b) cut forsterite surface with $\Theta_{\mathrm{i}}$ = 0.2, 0.3, 0.5, 0.6, 0.7, 0.8, 0.9 and 1.00 ML. In all cases $\nu$ = $1\times10^{13}$ s$^{-1}$ and $\beta$ = 0.3 K s$^{-1}$, $n=1$.}
\label{Fig:Simulation}
\end{figure}

In order to demonstrate the effect of the coverage dependent desorption energy on the thermal desorption, TPD traces were simulated by numerically integrating the Polanyi-Wigner equation and incorporating $E_{\mathrm{des}}(\Theta)$ for the initial relative coverages represented by the experimental traces in Figures~\ref{Fig:TPD_Cleaved} and ~\ref{Fig:TPD_Cut}. The experimental heating rate of 0.3 K s$^{-1}$ and a pre-exponential factor of $10^{13}$ s$^{-1}$ were used in performing the simulations. The fourth-order Runge-Kutta method was employed to solve the differential equations. The corresponding simulated desorption traces for NH$_{3}$ adsorbed on the cleaved and cut forsterite(010) surfaces are shown in Figures~\ref{Fig:Simulation}(a) and (b), respectively. The simulated TPD traces clearly demonstrate the difference in the distribution of higher adsorption energy sites on the two surfaces. As a result of the presence of a larger number of defects on the cut surface, the desorption traces extend to a significantly higher temperature of $T\sim 240$ K as shown in Figure~\ref{Fig:Simulation}(b), compared to $T\sim 220$ K for the cleaved surface as shown in Figure~\ref{Fig:Simulation}(a).

\section{Discussion}

Multilayer NH$_{3}$ desorption from both the cleaved and the cut crystal surfaces displays zero order kinetics where the leading edges are aligned as evident from Figures~\ref{Fig:TPD_Cleaved}(b) and \ref{Fig:TPD_Cut}(b). This suggests that the desorption kinetics for multilayer desorption, beyond the first ML, are insensitive to the underlying surface. Therefore, adsorbate-adsorbate interactions dominate over adsorbate-substrate interactions in this regime. The value of $E_{\mathrm{des}}=(25.8\pm 0.9)$ kJ mol$^{-1}$ is in good agreement with previous studies investigating the desorption of multilayer NH$_{3}$ ices deposited on a variety of substrates. A similar desorption energy of $E_{\mathrm{des}}=25.6$ kJ mol$^{-1}$ was obtained for the desorption of NH$_{3}$ from NH$_{3}$ ice by measuring the rate of ice sublimation as a function of temperature using IR absorption intensities \citep{Sandford1993}. Similarly, a desorption energy of $E_{\mathrm{des}}\sim25.5$ kJ mol$^{-1}$ was obtained for the desorption of pure thick NH$_{3}$ ice deposited on a potassium bromide (KBr) window \citep{Martin2014}. TPD of multilayer NH$_{3}$ from HOPG \citep{Bolina2005} yielded a multilayer desorption energy of $E_{\mathrm{des}}=(23.2\pm1.2)$ kJ mol$^{-1}$, in reasonable agreement with our value. However, these values are somewhat smaller than the sublimation energy value of 31.8 kJ mol$^{-1}$ for multilayer NH$_{3}$ ices obtained from a quartz crystal microbalance technique \citep{Luna2014}. This could be due to sublimation in this case occurring at a higher temperature of $T=112.5-113.5$ K compared to the desorption temperature range of $T=95-105$ K in the present case. In the work of \citet{Bolina2005}, the desorption of NH$_{3}$ from the HOPG surface was found to display a fractional desorption order of $n\sim0.25$, attributed to stronger hydrogen bonding between NH$_{3}$ molecules. In the present case we see a close alignment of leading edges in multilayer desorption and thus assumed zero order kinetics in this regime, consistent with the other NH$_{3}$ adsorption systems studied previously \citep{Kay1989, Szulczewski1998}.

For desorption from sub-ML coverages we see the appearance of a high-temperature tail for both surfaces, but with a marked difference between the two substrates as shown in Figures~\ref{Fig:TPD_Cleaved}(a) and \ref{Fig:TPD_Cut}(a). This tail is far more dominant and extends to higher temperatures for the rougher, cut surface. The appearance of such a high-temperature tail has previously been observed for the desorption of benzene from an amorphous SiO$_2$ substrate \citep{Thrower2009} and was attributed to desorption from a distribution of adsorption sites due to the rough surface. Recently, the desorption kinetics of CO$_2$ and H$_2$O from a crystalline olivine surface (011) with $10-15\%$ Fe$^{2+}$ have been investigated experimentally \citep{Smith2014}. A similar high-temperature desorption tail was attributed to the presence of a variety of adsorption sites arising from the different species present(O, Mg, Si, and Fe). In another recent study, the presence of high-energy sites influences the desorption kinetics, but for more volatile species adsorbed at lower temperature \citep{Collings2015}. TPD does not allow us to directly obtain information on the exact nature of the adsorption sites involved, but it is reasonable to consider that differences in adsorption strength between such different sites would also lead to the observed distribution of desorption energies in the present case. In the case of the cut forsterite surface we have an additional factor of surface roughness and the associated higher defect density which increases the contribution of high-energy adsorption sites, enhancing the high-temperature tail in the desorption traces. Similar extended desorption traces for the desorption of NH$_3$ from vapour-deposited amorphous silicate films have also been observed \citep{He2015}.

As the initial coverage of NH$_{3}$ is increased on both the cleaved and cut forsterite surfaces, the desorption gradually extends to lower temperatures until the multilayer peak starts to grow. We do not observe distinct desorption peaks associated with the different adsorption sites, but observe a continuous broad desorption feature. We can, however, clearly identify the transition between sub-ML and multilayer desorption. In a previous measurement of NH$_{3}$ desorption from quartz (0001), two peaks were assigned, with a broad $\alpha_1$ feature corresponding to adsorption at the surface defects at the lowest exposure and a lower temperature $\alpha_2$ feature which emerges with increasing coverage, saturating at the completion of the first ML \citep{Grecea2011}. In the present case, the presence of additional adsorption sites due to the presence of Mg atoms in the forsterite sample likely leads to an additional broadening of the desorption peak, which merges with the defect tail in the case of the cut forsterite sample. Contrary to the desorption behaviour observed on these silicate substrates,  NH$_3$ does not wet the graphite surface, leading to the absence of a distinct ML desorption peak as a result of the rather weak NH$_3$-graphite interaction \citep{Bolina2005}. Similar non-wetting behaviour was also observed for NH$_3$ adsorbed on polycrystalline Au and amorphous solid water substrates \citep{Collings2004}. This leads to the conclusion that NH$_3$ is bound more strongly to silicates than graphitic or water ice surfaces.

For the desorption of NH$_3$ from Ru(0001) \citep{Benndorf1983} the gradual shift of the ML desorption peak to lower temperature with increasing coverage was attributed to repulsive lateral interactions between neighbouring NH$_3$ dipoles which act to reduce the observed desorption energy. On Ru(0001), as for other metals such as Au(111) \citep{Kay1989}, NH$_3$ binds through the nitrogen lone pair electrons. Due to the close-packed structure of these metal surfaces and the preferred adsorbate orientation, lateral interactions are expected to play an important factor, whereas for forsterite there are different adsorption sites and at low coverage NH$_3$ molecules are further apart. Thus, although we cannot completely rule out some degree of repulsion at higher coverage, we would not expect that such interactions are a dominant factor in the significant shift of the desorption peak to lower temperature with increasing coverage at low coverage.

The shift to lower temperature could also be attributed to the pre-exponential factor being coverage dependent. In the framework of transition state theory (TST), the pre-exponential factor is related to the entropy change of desorption, \textit{i.e.} in going from the adsorbed state to the transition state. According to TST, the desorption rate is given by \citep{Laidler1940}

\begin{equation}
r_{\mathrm{des}} = \frac{k_{\mathrm{B}} T}{h} \frac {Q^\dagger}{Q} \exp \left[{-\frac {E_{\mathrm{des}}}{RT}}\right] \Theta ,
\label{eq:TST desorption rate eq}
\end{equation}

where the pre-exponential factor is given by

\begin{equation}
\nu = \frac{k_{\mathrm{B}} T}{h} \frac {Q^\dagger}{Q} ,
\label{eq:Pre-exponential eq}
\end{equation}

and $Q$ , $Q^\dagger$ are the partition functions in the adsorbed and gas-like transition states, respectively, $E_{\mathrm{des}}$ is the desorption energy, $T$ the surface temperature, $R$ the universal gas constant, $k_{\mathrm{B}}$ the Boltzmann constant and $h$ the Planck constant. The change in entropy, $\Delta S$ is related to the ratio of the partition functions in the two states. 

We can calculate two extreme limits for the pre-exponential factor. In the fully mobile lower limit, $\nu_{\mathrm{mobile}}$, we assume the adsorbate to have all possible degrees of freedom resulting in a minimum change in entropy in going from the adsorbed to the gas phase state. As it is only the rotational degrees of freedom that differ in the two states, this can be expressed in terms of one-dimensional and three-dimensional rotational partition functions in the adsorbed and transition states, respectively \citep{Gottfried2006,Thrower2013}:

\begin{equation}
\nu_{\mathrm{mobile}} = \frac {k_{\mathrm{B}} T}{h} \frac {q^\dagger_{\mathrm{rot,3D}}}{q_{\mathrm{rot,1D}}} = \left({\frac{k_{\mathrm{B}} T}{h}}\right)^2 \frac {1}{c\sqrt{B_A B_B}},
\label{eq:Fully mobile eq}
\end{equation}

where q$_{\mathrm{rot,1D}}^\dagger$ and q$_{\mathrm{rot,3D}}$ are the rotational partition functions in one and three dimensions, respectively. Rotational constants $B_A=B_B=9.44430$ cm$^{-1}$ have been obtained \citep{Herzberg1966}. This gives a value of $\nu_{\mathrm{mobile}}=1.53\times10^{13}$ s$^{-1}$ at a representative desorption temperature of 100 K. On the other hand, the immobile upper limit, $\nu_{\mathrm{immobile}}$, considers that an adsorbate has fully hindered degrees of freedom which results in a maximum change of entropy in going from the adsorbed state to the gas phase, as the molecule gains the translational degrees of freedom normal to the desorption coordinate:

\begin{equation}
\nu_{\mathrm{immobile}} = \frac {k_{\mathrm{B}} T}{h} q^\dagger_{\mathrm{trans,2D}} q^\dagger _{\mathrm{rot,3D}},
\label{eq:Immobile eq}
\end{equation}

where

\begin{equation}
q^\dagger_{\mathrm{trans,2D}} = \frac {2\pi m k_{\mathrm{B}} TA}{h^2}
\label{eq: Trans partition function eq}
\end{equation}

\begin{equation}
q^\dagger_{\mathrm{rot,3D}} = \frac{\pi^{1/2}}{\sigma} \left(\frac {k_{\mathrm{B}} T}{hc}\right)^{3/2}(B_A B_B B_C)^{-1/2}.
\label{eq:Rot partition function eq}
\end{equation}

$A$ is the surface area per adsorbed molecule, with a value $1.21\times 10^{-19}$ m$^2$ \citep{Bolina2005}. $\sigma$ denotes the symmetry factor of 3 for NH$_3$ molecules having a point group symmetry of C$_{3v}$. The rotational constant $B_C$ = 6.19600 cm$^{-1}$ has been obtained from \citet{Herzberg1966}. This yields a value of $\nu_{\mathrm{immobile}}=2.08\times10^{15}$ s$^{-1}$ at 100 K. Thus, the maximum change in pre-exponential factor that can be expected is around two orders of magnitude.

\begin{figure}
\includegraphics[width=\columnwidth]{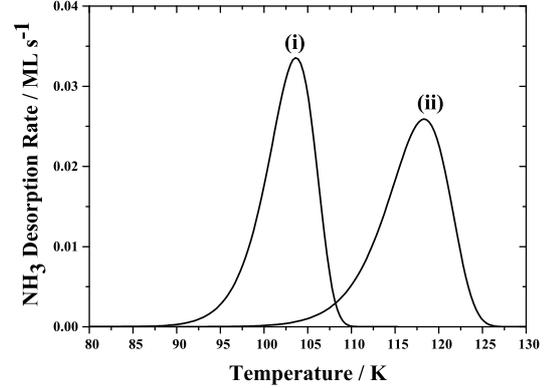}
\caption{Simulated TPD traces with NH$_3$ coverage $\Theta=0.8$ ML for the (i) immobile, $\nu_{\mathrm{immobile}}=2.08\times10^{15}$ s$^{-1}$ and (ii) fully mobile, $\nu_{\mathrm{mobile}}=1.53\times10^{13}$ s$^{-1}$ limiting case. In all cases a fixed desorption energy of $E_{\mathrm{des}}=32.3$ kJ mol$^{-1}$ was used along with $\beta=0.3$ K s$^{-1}$, and $n=1$.}
\label{Fig:Pre_exponential}
\end{figure}

\begin{figure*}
 \centering
		\subfloat{\includegraphics[width=0.4\textwidth]{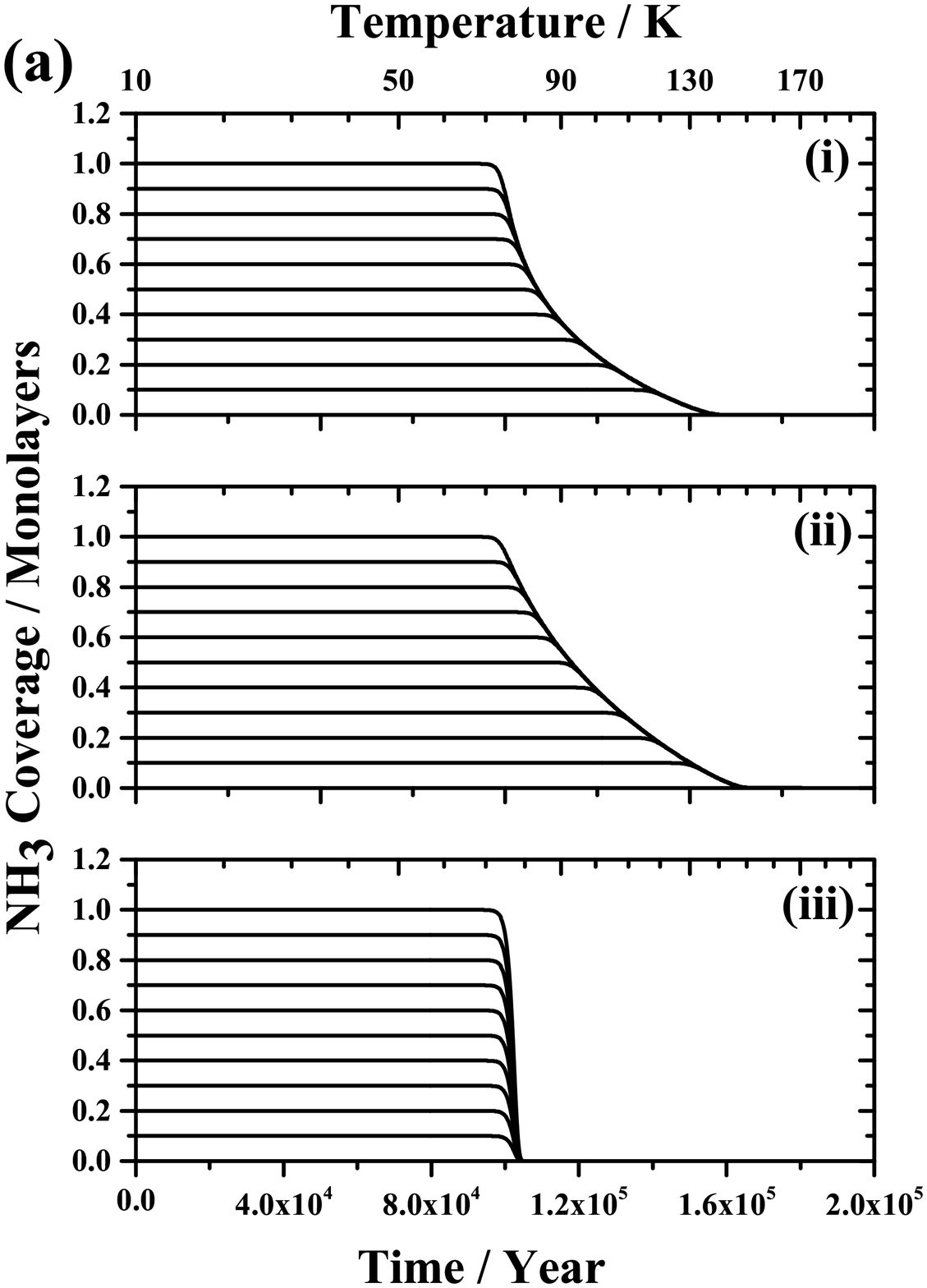}\label{Fig:Astrosim_cleaved}}
		\subfloat{\includegraphics[width=0.4\textwidth]{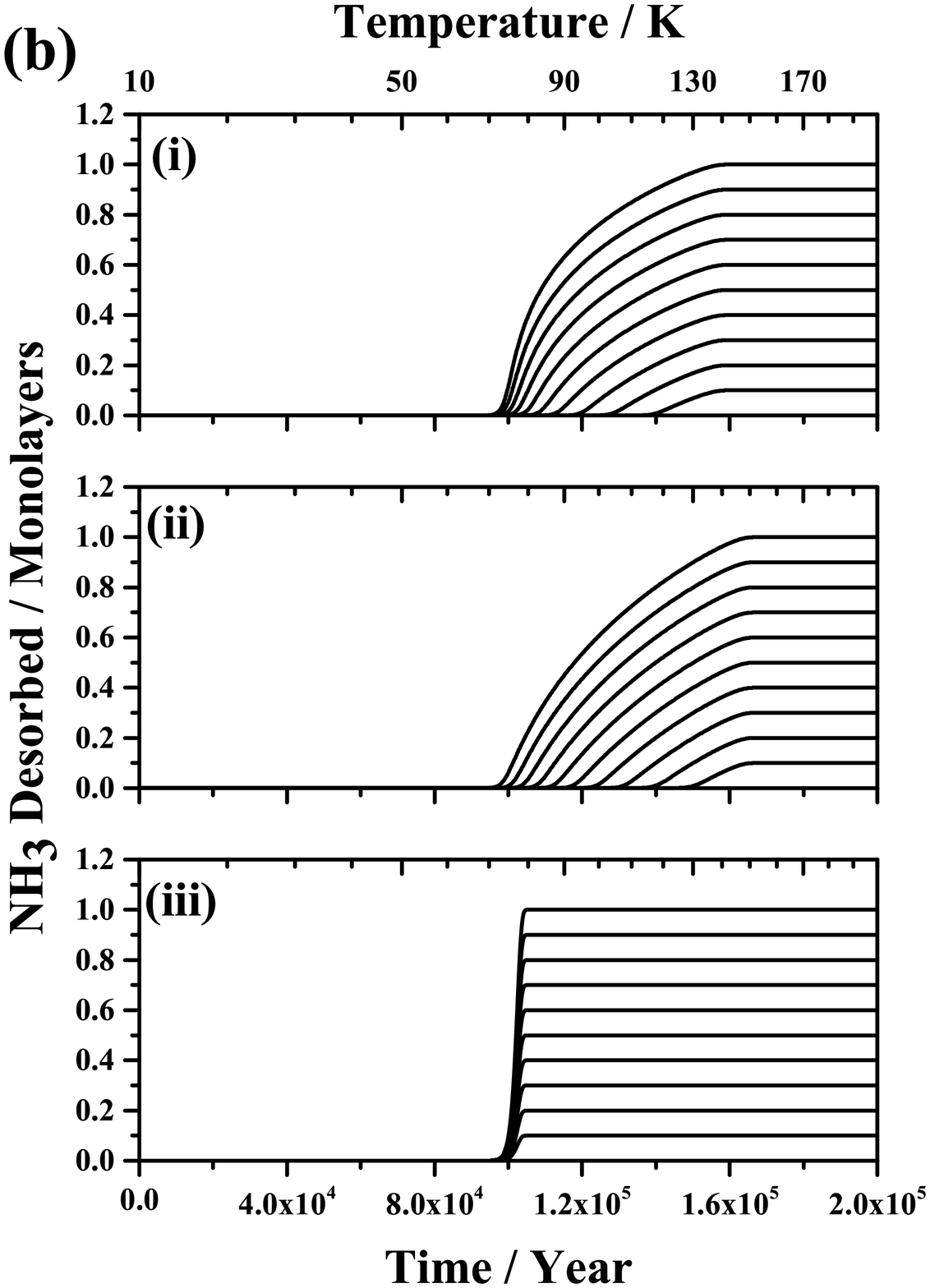}\label{Fig:Astrosim_cut}}
 \caption{Simulated (a) NH$_3$ coverage remaining and (b) NH$_3$ desorbed into the gas phase obtained using an astrophysically relevant non-linear heating profile (see text for details). The simulations incorporate the experimentally derived kinetic parameters obtained for (i) cleaved forsterite (010) and (ii) cut forsterite and are compared to (iii) the case of a constant desorption energy $E_{\mathrm{des}}$ = 27 kJ mol$^{-1}$.}
\label{Fig:Astrosim}
\end{figure*}

Coverage-dependent pre-exponential values have been obtained for the desorption of large molecules such as the polycyclic aromatic hydrocarbon coronene (C$_{24}$H$_{12}$; \citep{Thrower2013}) and long chain alkanes \citep{Campbell2012}. For large molecules, there is a significant change in entropy in going from an immobilized adsorbed state to a highly mobile gas-like transition state, leading to a significant change in $\nu$ with increasing coverage as the translational and rotational degrees of freedom become hindered in the dense adsorbed state. However, for small physisorbed species such as NH$_3$, the change in the pre-exponential factor with increasing coverage is relatively small and cannot account for the coverage dependent kinetics we observe experimentally. 

To further illustrate the effect of a changing pre-exponential on the desorption kinetics, we have simulated TPD traces for the two limiting values of $\nu$ for a fixed desorption energy. In order to obtain a representative value for $E_{\mathrm{des}}$ we have inserted the $\nu_{\mathrm{immobile}}$ limiting value into the Redhead expression and calculated the corresponding value for $E_{\mathrm{des}}$ based on the value of $T_\mathrm{max}$ for the cut forsterite surface with an initial coverage $\Theta_\mathrm{i}=0.8$, which yields $E_{\mathrm{des}}=32.3$ kJ mol$^{-1}$. The simulated TPD traces are plotted in Figure~\ref{Fig:Pre_exponential}. Trace (i) is obtained for $\Theta_{\mathrm{i}}=0.8$ ML, $\nu_{\mathrm{immobile}}=2.08\times10^{15}$ s$^{-1}$ and $E_{\mathrm{des}}=32.3$ kJ mol$^{-1}$ representing the immobile upper limit for $\nu$. The maximum desorption rate occurs at $T_{\mathrm{max}}\sim 103$ K. On the other hand, if the same kinetic parameters are used in the simulation, but with $\nu_{\mathrm{mobile}}=1.53\times10^{13}$ s$^{-1}$ representing the fully mobile lower limit, desorption trace (ii) is obtained. The maximum desorption rate now occurs at $T_{\mathrm{max}}\sim 118$ K. This shows that the maximum possible change in $\nu$ can only account for a rather small (\textit{ca.} 15 K) shift in desorption temperature and cannot explain the broad desorption tail that extends by many tens of K. This indicates that, although some change in $\nu$ with increasing coverage cannot be ruled out, the dominant effect on the desorption kinetics is that of an apparent coverage dependent desorption energy resulting from a distribution of adsorption sites. The choice of $E_{\mathrm{des}}$ in these simulations is rather arbitrary, and serves simply to demonstrate the maximum extent to which a change in $\nu$ alone can affect the observed desorption behaviour.

\section{Astrophysical Implications}

Laboratory kinetic data can be used to simulate thermal desorption on astrophysically relevant time-scales provided that appropriate heating rates are used. Heating rates in astronomical time-scales are much lower than laboratory based measurements. We have incorporated, in our simulations, an astrophysically relevant non-linear heating rate profile, as proposed for the warm-up phase in gas-grain chemical models \citep{Garrod2006}. Accordingly, we assume a temperature profile $T(t)=10+bt^2$, where $T$ is temperature in Kelvin at time $t$ in years and $b$ is a constant representing the heating rate. It was also proposed that the temperature increase from 10 to 200 K during warming-up and occurs at three different time-scales of $5\times10^4$, $2\times10^5$, and $1\times10^6$ yr in the case of high-, intermediate- and low-mass star formation, respectively. To examine the influence of the substrate on the desorption time-scale, we use the temperature profile for the case of intermediate mass stars with $b=4.75\times10^{-9}$ K yr$^{-2}$ \citep{Garrod2006}. The Polanyi-Wigner equation was numerically integrated, as before, incorporating the experimentally derived kinetic parameters and the non-linear heating profile. In order to understand the effect of the underlying substrate on the desorption time-scale of NH$_3$, the surface coverage and total amount of NH$_3$ desorbed into the gas phase are plotted as a function of time in Figures~\ref{Fig:Astrosim}(a) and \ref{Fig:Astrosim}(b), respectively. In both cases, we simulate the desorption of NH$_3$ employing the kinetic parameters for both the (i) cleaved and (ii) rough-cut surfaces as well as (iii) for a single desorption energy of $E_{\mathrm{des}}=27.0$ kJ mol$^{-1}$, which is close to that obtained for both surfaces at close to completion of the first ML. A single desorption energy would be expected in the case of desorption kinetics that are not coverage dependent for a surface that exhibits only one type of adsorption site. This allows us to examine the impact of both the presence of different adsorption sites and the presence of additional defects on the desorption time-scale.

We observe that the presence of higher adsorption energy sites will result in a return of species to the gas phase at much later times during the warm-up phase during star formation. From the simulations, it is evident that the presence of a coverage dependent desorption energy results in the extension of NH$_3$ desorption time-scale by about $5.9\times10^4$ yr compared to that obtained in the absence of a distribution of adsorption energies. In addition, the presence of a higher defect density on the cut forsterite additionally shifts the desorption time-scale by a further $7\times10^3$ yr compared to the cleaved forsterite surface. The dominant factor in determining the desorption time-scale is therefore the presence of a distribution of adsorption sites resulting from the chemical nature of the forsterite surface. We note that the lower desorption temperatures in these simulations compared to the laboratory measurements are a direct consequence of the significantly slower heating rate. Thus, on astrophysical time-scales, NH$_3$ desorption starts at \textit{ca.} 70 K as compared to 90 K for the laboratory measurements.

\section{Conclusions}
We have investigated the thermal desorption of pure NH$_3$ ice from cleaved and cut forsterite(010) surfaces and obtained the associated desorption kinetic parameters. This study provides experimental data for NH$_3$ desorption from two relevant model silicate surfaces that can be used in simulating thermal desorption under appropriate astrophysical conditions. Zero-order desorption kinetics are observed for multilayer desorption of NH$_3$, indicative of attractive adsorbate-adsorbate interactions between NH$_3$ molecules. The obtained multilayer desorption energy of $E_{\mathrm{des}}=(25.8\pm 0.9$) kJ mol$^{-1}$ is in good agreement with previous studies and is insensitive to the nature of the underlying substrate. At sub-ML coverages, where the adsorbate-substrate interaction is expected to play a larger role, the high-temperature tail observed in the desorption profiles clearly demonstrates the influence of the grain surface chemistry and morphology on the desorption kinetics. This broad desorption feature can be attributed to a sequential filling of a broad distribution of adsorption sites on the forsterite (010) surface which result from the surface chemical structure of the crystal. The presence of a larger number of defects such as steps and kinks on the rougher, cut forsterite surface compared to the cleaved forsterite surface leads to an even broader distribution of adsorption energies. This is exemplified by \textit{ca.} 77\% of desorption sites displaying adsorption energies greater than 30 kJ mol$^{-1}$ for the cut forsterite surface compared to \textit{ca.} 58\% for cleaved forsterite. Kinetic simulations incorporating a coverage dependent desorption energy reproduce the experimentally observed behaviour. Simulated thermal desorption profiles incorporating astrophysically relevant heating rates reveal how these differences in the underlying substrate strongly influences the time-scale on which NH$_3$ is returned to the gas phase. Compared to a simple first order desorption characterized by a single desorption energy of 27 kJ mol$^{-1}$, the presence of a distribution of desorption energies leads to an increase in desorption time of the order of $5.9\times10^4$ yr. The presence of a higher defect density on a rough silicate grain surface, as modelled by the cut forsterite sample, results in desorption extending to somewhat later times of around $7\times10^3$ yr compared to the single crystal surface. Our study has used crystalline forsterite as a suitable dust grain model with which to study the desorption kinetics of small molecules. This study emphasized that the morphology of dust grains has a strong influence on the desorption of adsorbed species by providing a range of binding sites and thus extending the desorption time-scale significantly.

\section*{Acknowledgements}

We acknowledge financial support from the European Commission's 7th Framework Programme through the 'LASSIE' ITN under Grant Agreement no. 238258, the BMBF through FSP 301 'FLASH' and the NRW International Graduate School of Chemistry, M{\"u}nster. We thank Dr. F. Schappacher and Dr. F. Winter (MEET battery research centre of University of M{\"u}nster) for performing the XRD measurements. We would also like to thank H. Mutschke (Friedrich-Schiller-University Jena) for IR reflectance measurements on the forsterite crystal and Dr. H. King (University of Utrecht) for helpful discussion regarding the surface crystal structure of forsterite.




\bibliographystyle{mnras}
\bibliography{references} 








\bsp	
\label{lastpage}
\end{document}